\documentclass[%
 reprint,
 amsmath,amssymb,
 aps,
]{revtex4-1}

\usepackage{graphicx}
\usepackage{dcolumn}
\usepackage{bm}
\usepackage{dsfont}

\usepackage{color}
\usepackage{ulem}

\begin{document}

\preprint{APS/123-QED}

\title{Note on the gluino condensate in supersymmetric Yang Mills theory}

\author{Renata Jora
	$^{\it \bf a}$~\footnote[1]{Email:
		rjora@theory.nipne.ro}}

\affiliation{$^{\bf \it a}$ National Institute of Physics and Nuclear Engineering PO Box MG-6, Bucharest-Magurele, Romania}

\begin{abstract}

We calculate gluino condensate for supersymmetric Yang Mills theory on general grounds without making any assumption with regard to the weak or strong regime of the theory. Our result coincides with that obtained from an instanton superpotential when the theory is weakly interacting. The method also determines the corresponding partition function.

\end{abstract}
\maketitle

Supersymmetric gauge theories have long been a laboratory not only for testing beyond standard model paradigms but also for their phase structure that might be relevant for the more mundane counterpart QCD. In studying the nonperturbative behavior of a gauge theory of utmost relevance is the vacuum structure of the theory.  This is strongly related to the underlying internal or global symmetries.  It is rarely easy to determine the various strong gauge condensates from first principles. However there are instances where the analytical structure of the theory or powerful symmetries may provide straightforward answers. One of these cases is supersymmetric Yang Mills theory with the gluino condensate calculated in the weakly interacting regime  \cite{Pouliot}, \cite{Shifman}, \cite{Morozov} as:
\begin{eqnarray}
\langle\lambda^a\lambda^a\rangle=32\pi^2\Lambda^3.
\label{first554}
\end{eqnarray}
Here $\Lambda$ is the intrinsic scale of supersymmetric Yang Mills corresponding to,
\begin{eqnarray}
\frac{8\pi^2}{Ng_h^2}=\ln(\frac{\mu}{\Lambda}),
\label{secres65774}
\end{eqnarray}
where $\frac{1}{g_h^2}$ is the holomorphic coupling constant,
\begin{eqnarray}
\frac{1}{g_h^2}=\frac{1}{g^2}+i\frac{\theta}{8\pi^2}.
\label{def45553}
\end{eqnarray}

In the strong coupling regime of supersymmetric Yang Mills the situation is more complex \cite{Pouliot}, \cite{Novikov}, \cite{Amati} and the results do not coincide with that in the weak regime even in the large N limit:
\begin{eqnarray}
\langle\lambda^a\lambda^a\rangle_{strong}\approx\frac{1}{N} \langle\lambda^a\lambda^a\rangle_{weak}.
\label{tsronres664553}
\end{eqnarray}

In this work we reiterate the calculation of the gluino condensate on general grounds and using  a different method without specifying the particular regime of the theory.

The Lagrangian for supersymmetric Yang Mills in terms of the holomorphic coupling has the expression:
\begin{eqnarray}
{\cal L}_h=\frac{1}{16}\int d^2\theta \frac{1}{g_h^2} W^a(V_h)W^a(V_h)+h.c.,
\label{holo999}
\end{eqnarray}
where $W^a_{\alpha}(V_h)=-\frac{1}{4}\bar{D}^2\exp[-2V_h]D_{\alpha}\exp[2V_h]$.

One can obtain a Lagrangian in terms of the canonical coupling  by making the change of variable $V_h=g_cV_c$ \cite{Murayama}.  Then the beta function in the canonical coupling is the NSVZ beta function \cite{NSVZ}:
\begin{eqnarray}
\beta(g_c)=-\frac{g_c^3}{16\pi^2}\frac{3N}{1-N\frac{g^2_c}{8\pi^2}}.
\label{betaufnc665}
\end{eqnarray}

The holomorphic and canonical coupling constants are related by \cite{Murayama}:
\begin{eqnarray}
\frac{1}{g_h^2}=\frac{1}{g_c^2}+\frac{N}{8\pi^2}\ln(g_c^2).
\label{connection5664}
\end{eqnarray}

If $Z$ is the partition function for the supersymmetric Yang Mills theory,
\begin{eqnarray}
Z=\int d V_h\exp[i\int d^4 x{\cal L}],
\label{partfunction776566}
\end{eqnarray}
then one can promote the holomorphic coupling to a background chiral superfield and has:
\begin{eqnarray}
\frac{\delta \ln Z}{\delta \frac{1}{g_h^2}}=\langle W^a_{\alpha}W^{a\alpha}+h.c.\rangle=\langle \lambda^a_{\alpha}\lambda^{a\alpha}+h.c.\rangle,
\label{res664553}
\end{eqnarray}
where the right hand side of the equation corresponds to the gluino condensate and the derivative is in a functional sense.  Then one can write:
\begin{eqnarray}
\frac{d \ln Z}{d \frac{1}{g_h^2}}=\int d^4 x d^2 \eta [-\frac{1}{4}\langle W^a_{\alpha}W^{a\alpha}\rangle].
\label{newexpr7775665}
\end{eqnarray}
Here the derivative is in the ordinary sense and $\eta_{\alpha}$ are the  Grassmann variables. Noting that on dimensional reasons and because as a Green function the quantity  $\langle W^a_{\alpha}W^{a\alpha}\rangle$ is renormalization scheme independent it can have only the following expression $\langle W^a_{\alpha}W^{a\alpha}\rangle= b\Lambda^3$, where $\Lambda$ is the characteristic scale of super Yang Mills and $b$ is a constant to be determined. Since $\Lambda^3=\mu^3\exp[-\frac{8\pi^2}{Ng_h^2}]$ one can integrate Eq. (\ref{newexpr7775665}) to obtain:
\begin{eqnarray}
\ln Z=\frac{N}{32\pi^2}\int d^4x d^2\eta b \Lambda^3 +{\rm const}.
\label{res554yu}
\end{eqnarray}
Here the constant on the second term of the right hand of the Eq. (\ref{res554yu})  can be calibrated by asking $\ln Z=0$ for $g_h=0$ which leads to ${\rm const}=0$ (Note that even in the absence of this calibration our calculations although become slightly more complicated lead to the same final result for the gluino condensate).
Alternatively one may write:
\begin{eqnarray}
\ln Z=c\exp[-\frac{8\pi^2}{Ng_h^2}]=c\frac{\Lambda^3}{\mu^3},
\label{alt466577}
\end{eqnarray}
where $c$ is a constant to be determined.

Now consider the action and the partition function in the background gauge field of an instanton with $n=1$. The  partition function in this case has the expression:
\begin{eqnarray}
&&Z={\rm const}\exp \Bigg[-\frac{1}{4g_h^2}\int d^4  xd^2 \eta W^a_{\alpha}(B_h)W^{a\alpha}(B_h)+
\nonumber\\
&&{\rm higher\,\,order\,\,terms}\Bigg],
\label{partfunc6645537}
\end{eqnarray}
where $B_h$ is the background gauge superfield.

The higher order terms in the background gauge field would necessarily have higher orders of factors $\frac{1}{g_h^2}$. Then we consider the scale anomaly associated to the partition function in Eq. (\ref{partfunc6645537}):
\begin{eqnarray}
&&\frac{d\ln Z}{d\ln(\mu)}=
\nonumber\\
&&-\frac{3N}{32\pi^2}\int d^4  xd^2 \eta W^a_{\alpha}(B_h)W^{a\alpha}(B_h)+...,
\label{resimpot77675}
\end{eqnarray}
where the dots stand for higher order terms. However if one considers the scale anomaly taken at $\mu=\Lambda$ then $\frac{1}{g_h^2}=\infty$ and consequently the higher order terms which contain higher powers of $\frac{1}{g_h^2}$ will have besides the constant expression associated to the beta function factors of $\frac{1}{g_h^2}$ and thus will lead to zero. For a standard normalization of the instanton:
\begin{eqnarray}
\int d^4  xd^2 \eta W^a_{\alpha}(B_h)W^{a\alpha}(B_h)=1,
\label{nomr65774}
\end{eqnarray}
one then obtains:
\begin{eqnarray}
\frac{d\ln Z}{d\ln(\mu)}|_{\mu=\Lambda}=-3N.
\label{imprt566477}
\end{eqnarray}

Now we consider the same anomaly associated to Eq. (\ref{alt466577}):
\begin{eqnarray}
\frac{d\ln Z}{d\ln(\mu)}|_{\mu=\Lambda}=-3c.
\label{secr54663}
\end{eqnarray}

Comparing Eqs. (\ref{imprt566477}) and (\ref{secr54663}) leads to the determination of the constant $c=N$ such that,
\begin{eqnarray}
\ln Z_{\mu=\Lambda}=N.
\label{exp7576}
\end{eqnarray}

Next we start form Eq. (\ref{res554yu}) to write:
\begin{eqnarray}
&&\ln Z=\frac{N}{32\pi^2}\int d^4x d^2\eta b \Lambda^3=
\nonumber\\
&&\frac{N}{32\pi^2}\int d^4x d^2\eta b \mu^3\times
\nonumber\\
&&\exp \Bigg[\frac{-8\pi^2}{N}[\frac{1}{g_h^2}+\frac{N}{8\pi^2}(i\mu x+i\mu\eta)]\Bigg]\times
\nonumber\\
&&\exp[-i\mu x-i\mu\eta].
\label{res76885775}
\end{eqnarray}
We make the change of variables $\mu x=x'$ and $\mu\eta=\eta'$ which yields:
\begin{eqnarray}
&&\ln Z=\frac{N}{32\pi^2}\int d^4x' d^2\eta' b \times
\nonumber\\
&&\exp\Bigg[\frac{-8\pi^2}{N}(\frac{1}{g_h^2}+\frac{N}{8\pi^2}(i x'+i\eta'))\Bigg]\times
\nonumber\\
&&\exp[-i x'-i\eta'].
\label{secondeq664553}
\end{eqnarray}
We further make a change of variable of the chiral superfield $\frac{1}{g_h^2}\rightarrow \frac{1}{g_h^2}+\frac{N}{8\pi^2}(i x'+i\eta')$. Then the expression in Eq. (\ref{secondeq664553}) is the Fourier transform of the chiral superfield (where we take directly $\mu=\Lambda$):
\begin{eqnarray}
\ln Z_{\mu=\Lambda}=\frac{N}{32\pi^2}b\exp[-\frac{8\pi^2}{g_h^2}](\frac{p}{\Lambda}=1)=\frac{N}{32\pi^2}b.
\label{finalres664554}
\end{eqnarray}

From Eqs. (\ref{exp7576}) and (\ref{finalres664554}) one then determines:
\begin{eqnarray}
&&b=32\pi^2
\nonumber\\
&&\langle \lambda^a_{\alpha}\lambda^{a\alpha}+h.c.\rangle=32\pi^2 \Lambda^3.
\label{finalres775664}
\end{eqnarray}

In this work we calculated the gluino condensate for supersymmetric Yang Mills theory without making any assumption with regard to the regime of the theory. Our estimate agrees with that obtained in the weak interaction limit \cite{Pouliot}, \cite{Shifman}, \cite{Morozov} in the absence of the premises induced in the computation for that case.  The method employed here may offer clues in the behavior of supersymmetric Yang Mills theory both in the strong and weak regime. As a byproduct we also determined the partition function.

\end{document}